\documentclass[10pt]{article}

\usepackage[english]{babel}

\usepackage[letterpaper,top=2cm,bottom=2cm,left=3cm,right=3cm,marginparwidth=1.75cm]{geometry}

\usepackage{amsmath}
\usepackage{graphicx}
\usepackage[colorlinks=true, allcolors=blue]{hyperref}

\title{\bf A Machine Learning system to monitor student progress in educational institutes}

\author{{Bibhuprasad Mahakud\textsuperscript{1}$^\ast$, Ipsit Panda\textsuperscript{1}, Souvik Maity\textsuperscript{1}},\\
{Arpita Sahoo\textsuperscript{1}, Reeta Sharma\textsuperscript{1}, Bibhuti Parida\textsuperscript{2}}\\
\textsuperscript{1}\href {https://diracai.com}{diracAI Private Limited, Bhubaneswar, India}\\
\textsuperscript{2}\href {https://www.amity.edu/}{Dept. of Applied Physics, AIAS, Amity University, Noida, India}\\
\href{$^\ast$bibhuprasad.mahakud@diracai.com}{$^\ast$bibhuprasad.mahakud@diracai.com}
}

\begin{document}
\maketitle

\begin{abstract}
In order to track and comprehend students' academic achievement, both private and public educational institutions devote a significant amount of resources and labour. One of the difficult issues that institutes deal with on a regular basis is understanding students' exam shortcomings. The performance of a student is influenced by a variety of factors, including attendance, attentiveness in class, understanding of concepts taught, the teacher's ability to deliver the material effectively, timely completion of home assignments, and the concern of parents and teachers for guiding the student through the learning process. We propose a data-driven approach that makes use of  Machine Learning (ML) techniques to generate a classifier called credit score that helps to comprehend students' learning journeys and identify activities that lead to subpar performances. This would make it easier for educators and institute management to create guidelines for system development to increase productivity. The proposal to use credit score as progress indicator is well suited to be used in a Learning Management System (LMS). In this article, we demonstrate the proof-of-the-concept under simplified assumptions using simulated data. 
\end{abstract}


\section{Introduction}

One of the primary priorities of every educational institution is keeping track of students' performance in their coursework. A instructor or the institute's director wishes to keep track of and comprehend students' exam performance.
Each learner encounters several challenges that impede their development. These problems can be caused by both internal and external influences. A student's bad performance may be caused by irresponsibility, such as failing to show up for class, becoming disengaged during class, or failing to turn in all of their homework on time, among other things. The student's growth and comprehension of the subject matter, however, may be hampered by outside variables as well. For instance, if the teacher is ineffective at delivering the material and clearly illuminating concepts, the pupils may perform poorly. Additionally, each student may present a unique use case in the classroom with relation to their performances, making it challenging for the institute's teacher to keep track of, monitor, and take timely corrective action to resolve these issues on a daily basis. When the teacher-to-student ratio is exceedingly low, this issue becomes labor-intensive and difficult. When the teacher-to-student ratio is very high, it might be possible for the instructor to manually watch each student to identify their shortcomings and bad habits and to establish a plan of action to address issues. However, at large institutions, it can be difficult for teachers to keep track of hundreds of pupils at once or for the government to keep tabs on the overall performance of numerous schools.

One approach to solving this issue would be to use a learning management system (LMS) \cite{klobas2010role} to access and store student curricular activity data across time, then train a machine learning (ML)~\cite{carbonell1983overview}  algorithm  to understand all of these activities and how they relate to student performance. Once trained, the classifier can offer insightful information about a student's performance and behaviour. The teacher would be able to identify the student's weaknesses that would have a negative impact on performance since it would be able to foresee how a student's mistake would affect performance. In situations when there is a low teacher to student ratio, our suggested approach will be more appropriate to address the issue. Additionally, it will make the job of the owner of the institute, who oversees thousands of pupils, easier.

To demonstrate the issue and outline the solutions that an ML system can offer in each scenario, we show a few use cases.

\subsection{Use case (i)}

Despite being on time and attending all classes, a student performs poorly on an exam. A machine learning system (MLS)~\cite{ke1988mls} that has been taught can identify the cause of subpar performance. This could be for one or more of the reasons listed below.

\begin{itemize}
\item The student may be copying all homework assignments and failing to pay attention in class.
\item It is possible that the student isn't solving enough problems aside from the assigned homework or assignments.
\item Before exams, for example, the student might not have revised the concepts.

\end{itemize}

\subsection{Use case (ii)}
A student does not comprehend what the instructor is teaching in class. A machine learning system that has been taught can determine the cause, which may be one or more of the following.

\begin{itemize}
  \item By not completing the prerequisites, the student might not be arriving prepared to the class.
  \item The student might not stay attentive throughout the lesson.
  \item It is possible that the instructor is not explaining the material clearly and effectively, etc.
\end{itemize}

When there are multiple parameters, the MLS can determine which one has a greater contribution.

\subsection{Use case (iii)}

The director of the institute wants to know if everyone completes their job as they are expected to. A ML system can identify how that will affect the performance if a few professors and students fail to finish their assignments on time. 

\subsection{Use case (iv)}

The institute director or teacher wishes to track a number that should indicate how well his institute or class is doing on a daily basis across a number of metrics involving hundreds of pupils. Exam results provide insight into an institution's performance, but they cannot be used as a daily gauge of how well daily operations are doing. However, a grade based on the ML classifier~\cite{malviya2020machine} can forecast how a student's performance would be impacted by daily events.

Investigating the following would be necessary in order to create an effective ML classifier.

\begin{itemize}
  \item All relevant parameters that are correlated with performance. 
  \item A sufficient amount of actual data to be randomly collected to reduce the discrepancy between the sample and parent distribution.

\end{itemize}

To demonstrate the concept's validity, we employed simulated data under a few generalised assumptions and a small number of intuition-based input parameters in the context of this study. However, the same idea might be modified and expanded upon for real-world situations. 

Regression has been used in the current scope to demonstrate the concept, however other machine learning approaches may also be used depending on the complexity of data.

\section{Input Parameters and Simulated data}

The actions that have an impact on a student's success on a particular exam are referred to in this note as input parameters. These include things like student presence in class, attentiveness, in-class comprehension, etc.

The input parameters used to train the ML algorithm are crucial in determining the algorithm's performance. The number of input parameters can be very large in a real-world setting. However, in this instance, we take into account a few parameters to demonstrate the proof-of-the-concept. The same could be enhanced by adding more parameters.

\subsection{Attendance}
One important characteristic that is thought to be strongly connected with exam success is regular or all-class attendance.  To understand the nature of attendance distribution, the binomial probability of attending a class (for a student) is taken to be a constant for all students. The scenario is considered with a coin-toss experiment~\cite{diaconis2007dynamical} which is Binomial~\cite{von2001binomial}  in nature. For unbiased and identical coins the probability of getting a head in a toss-experiment is described by a binomial. Under the condition of large number of trials ({\it n}), such that {\it n} times binomial probability({\it p}) remains constant, binomial converges to poisson. The mean of poisson is represented by $\mu$ ( {\it np} ). The poisson probability is given by:

\begin{equation}
P(n, \mu) = {\frac{\mu^{n}e^{-\mu}}{n!}}
\end{equation}

We take the attendance distribution to be a normal distribution assuming $\mu$ is sufficiently higher where poisson approximates a gaussian. The attendance distribution is shown in Fig. 1 (top left).

\begin{figure}[h!]
  \includegraphics[scale=0.4]{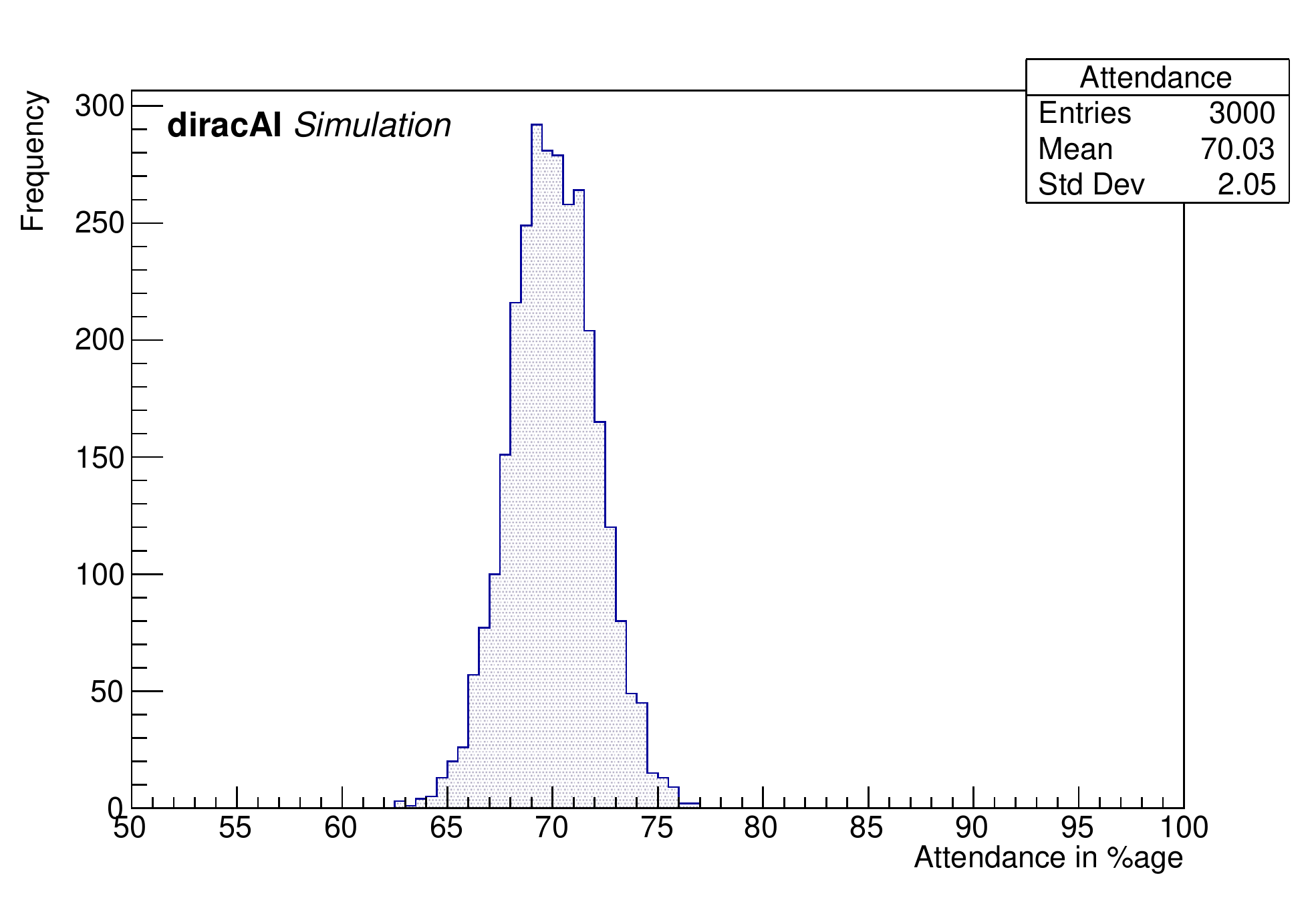}
  \includegraphics[scale=0.4]{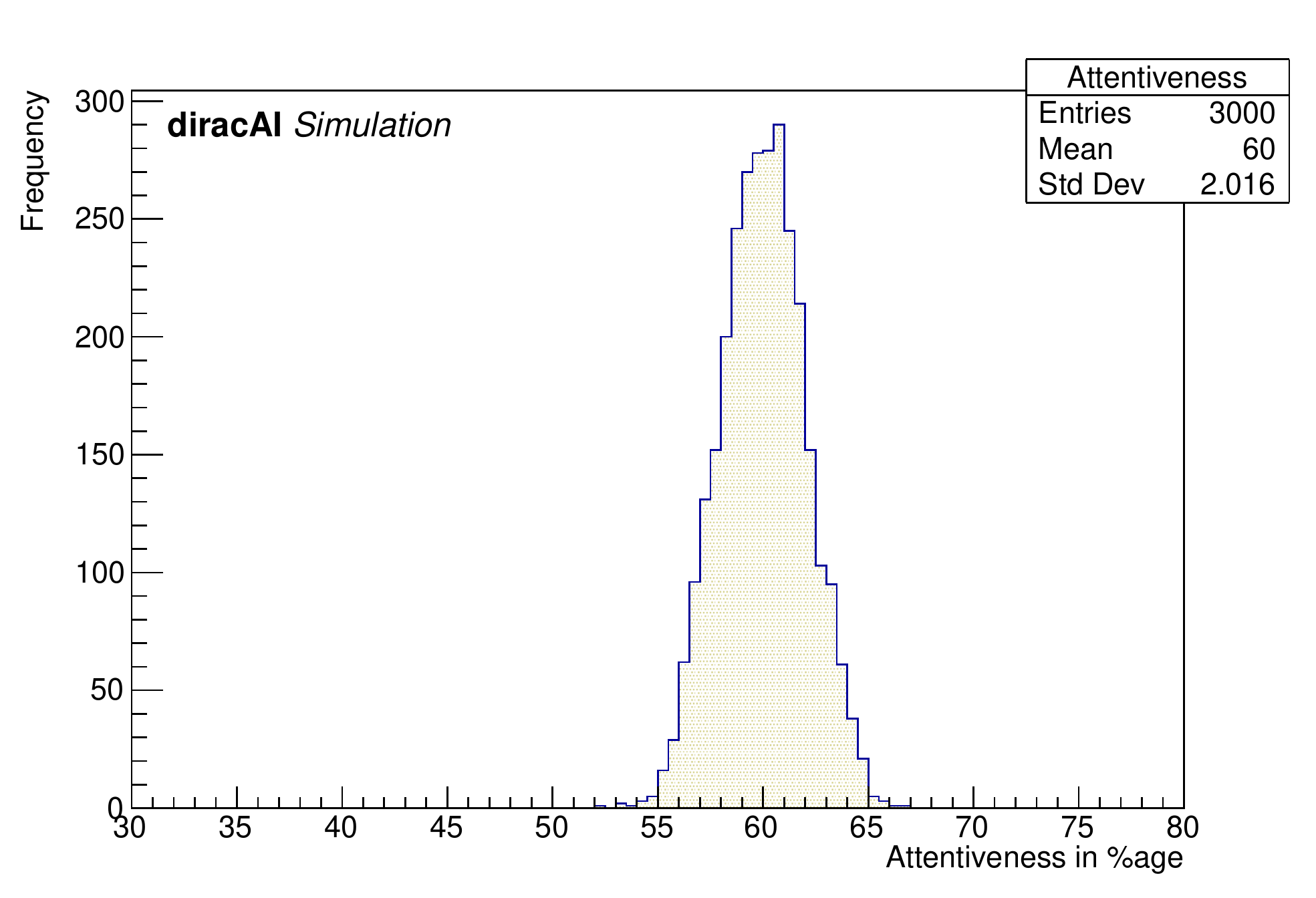}\\
  \includegraphics[scale=0.4]{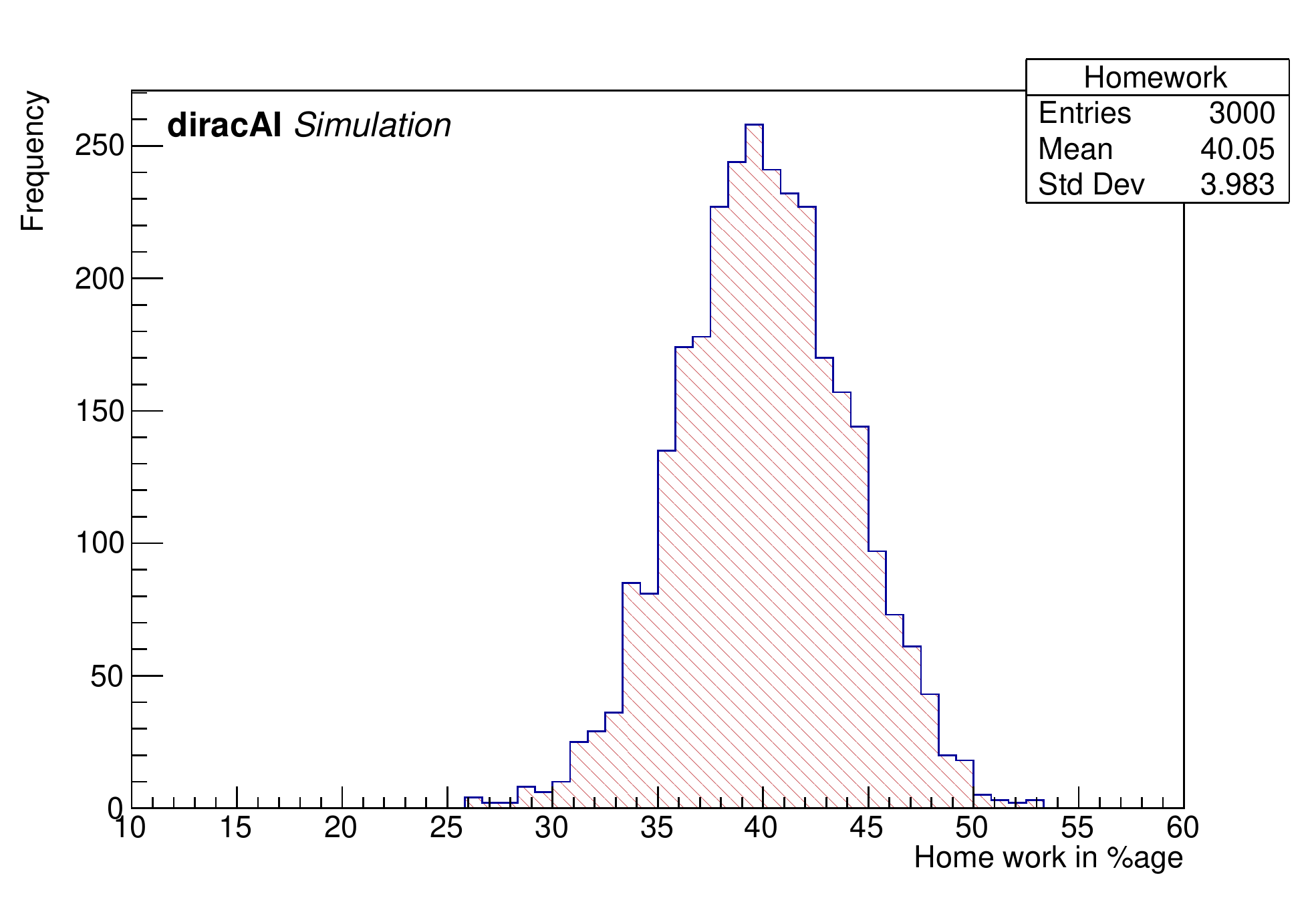}
  \includegraphics[scale=0.4]{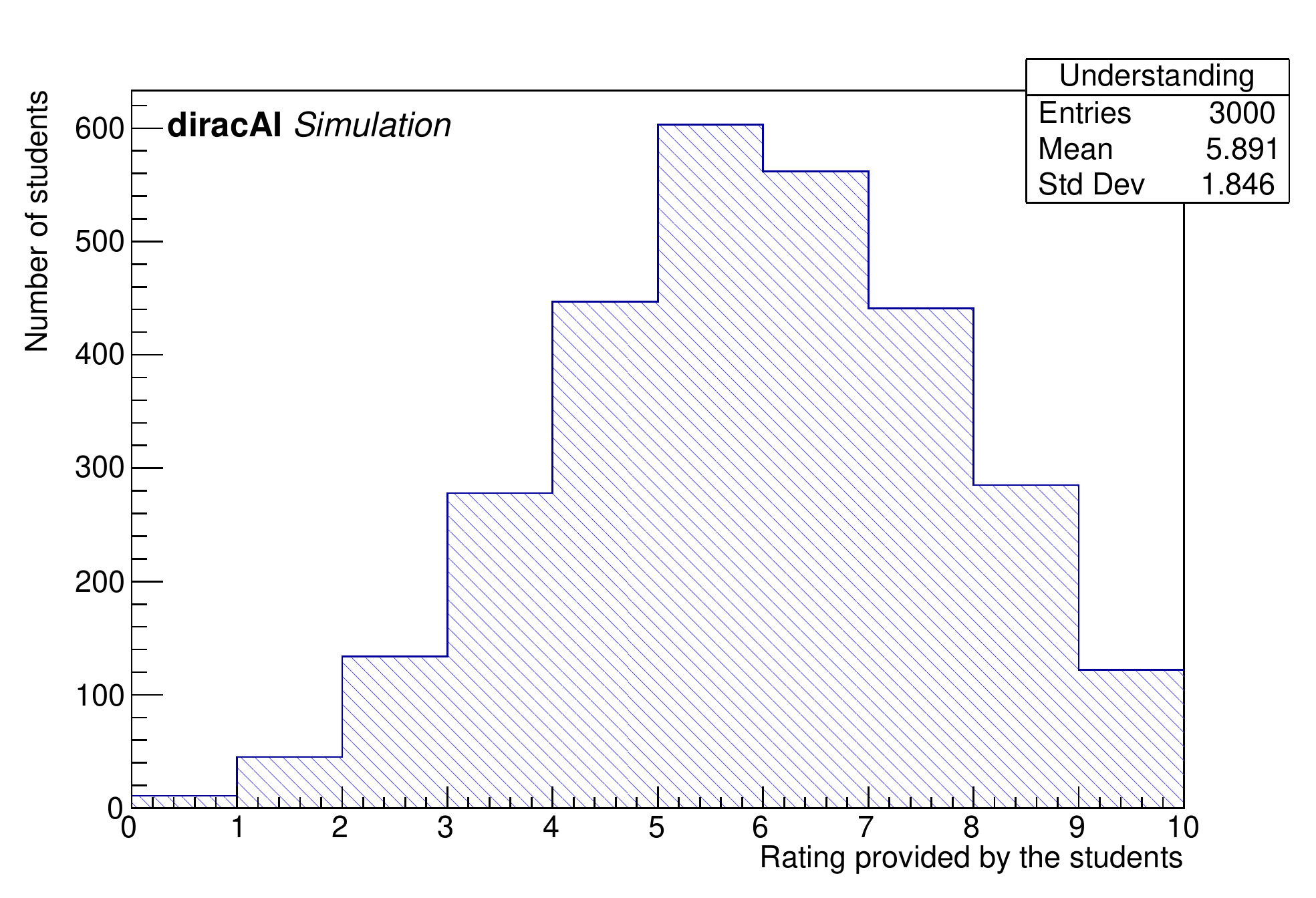}
  \caption{Attendance distribution. }
  \label{fig:birds1}
\end{figure}

\subsection{Attentiveness in class}

One would anticipate that a student would perform better if they remained focused in class. Remaining inattentive could hamper performance in several ways.  This is one of the crucial study parameters, in our opinion. Through the video chat programme, attention during the online classes may be monitored to some level. When a student reacts quickly to the pop-up window, the system can automatically inform the teacher in the class that they are aware. Additionally, a teacher may use a "poll-app" to ask students relatively basic questions that would validate their attention during class. The attentiveness distribution is shown in Fig.1 (top right).

\subsection{Homework}
One of the most important aspects of performance is doing homework on time. A student is more likely to comprehend the material in the forthcoming class and score better on exams if they turn in their homework on time. There are numerous forms of homework, which can be further divided into subcategories. In this instance, we treat it as a single category and suppose that homework is a brief assignment that teachers give out at the end of each lesson or occasionally throughout the semester.

 The distribution of homework is thought to occur more frequently nevertheless, following the teaching of each smaller unit of a subject. A course would have a finite amount of total homework. The teacher can then determine what percentage of the overall amount of homework has been completed and turned in. We provide simulated distribution of the fraction of homework done by the students in a course. According to Fig. 1 (bottom left), students typically finish only $70\%$ of the total amount of homework assigned by the teacher.

\subsection{Understanding in the class}

A significant portion of a student's time is spent in class. Understanding the information presented in class would have a significant impact on performance. In the classroom, understanding depends on a number of variables. These could include prerequisites, attentiveness, or the teacher's ability to effectively present the content. We have treated this as a separate category for this reason. The user rating could be used to gather comprehension data. A user (student) rating system with ten distinct categories has been created. The simulated rating  provided by students is displayed in Fig. 1's bottom right corner.

\subsection{ Previous Exam Performance }

Due to the conceptual clarity of fundamental concepts, a student who has a history of poor performance in past courses, years, or exams is likely to do poorly. The results of earlier tests might be used to evaluate this. For instance, a student in class 12 might be judged based on their performance in class 11. His or her performance in the current course may be tied to the score from the prior year. We believe that this is a crucial parameter to take into account while creating ML models. Below is a typical exam score distribution.

\begin{figure}[h!]
  \centering
  \includegraphics[scale=0.35]{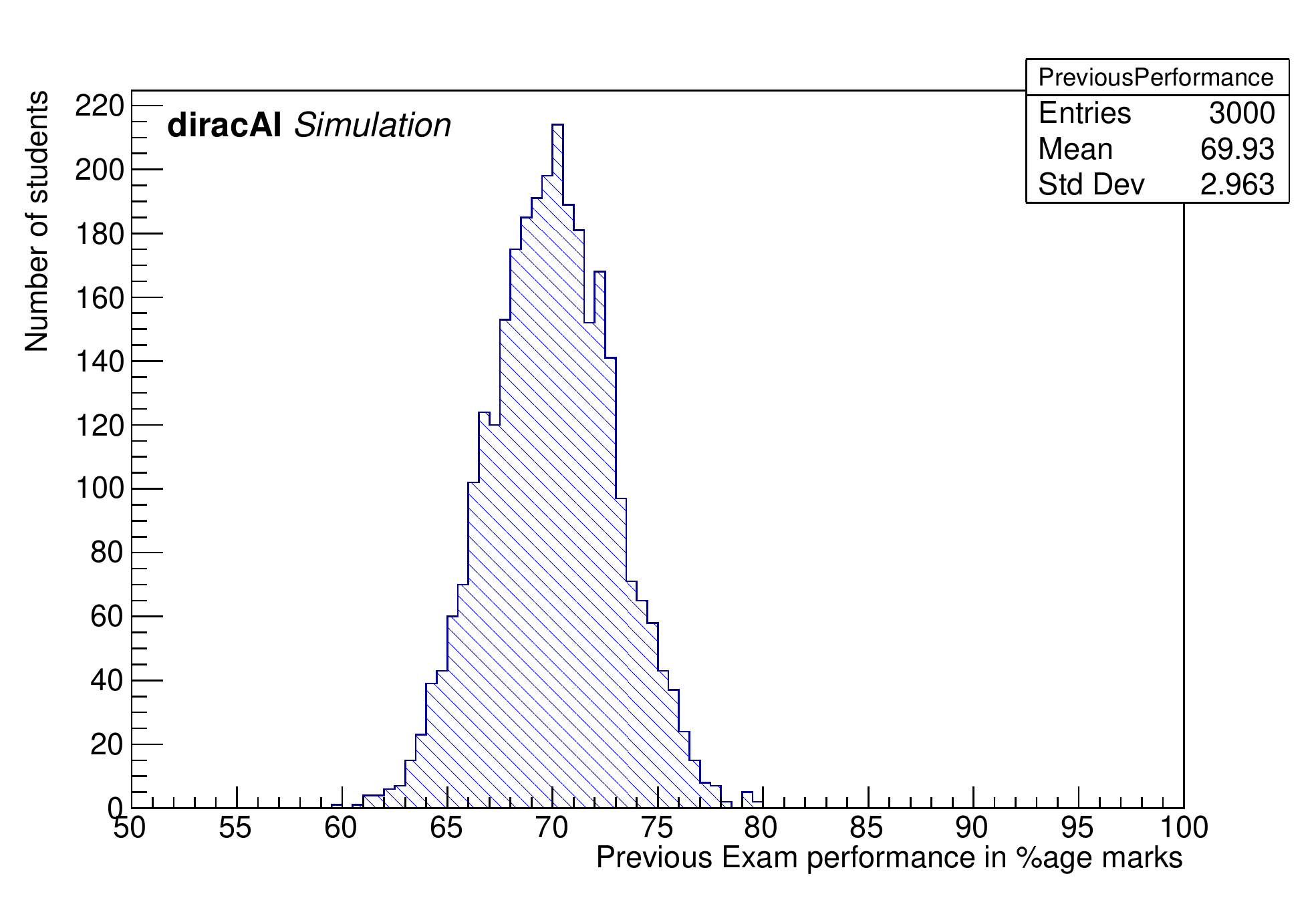}
  \caption{Previous exam performance }
  \label{fig:birds5}
\end{figure}

\section{ Performance Data Generation}

Performance data is obtained by taking a linear combination of all the above mentioned parameters and adding additional uncertainty on top of it. For the purposes of data production, the following linear parametrization is taken into consideration.

\begin{equation}
{\rm Performance} = {c0+ c1*X^{1}+ c2*X^{2}+ c3*X^{3}+ c4*X^{4}+ c5*X^{5}}
\end{equation}
where $X^{1}$,$X^{2}$, $X^{3}$, $X^{4}$ and $X^{5}$  are attendance, attentiveness, homework, understanding and previous performance respectively. c1, c2, c3, c4 and c5 are the weight factors assigned to each of them respectively. c0 is a constant. A table of data containing five inputs and one output is prepared. Table 1. shows a typical 5 set of data points. Simulated exam performance data is shown in Fig. 3.
 
\begin{table}
	\caption{Data points}
	\centering
	\begin{tabular}{llllll}
	\hline
%
		Attendance     & Attentiveness  & Homework & Understanding & Previous Performance   & Performance  \\
\hline
		67.9\%   & 59.9\%  &  30.6\%  &  9 & 67.4  &  72.9 \\
		73.5\%   & 63.2\%  &  36.3\%  &  3 & 68.3  &  79.4 \\
		71.6\%   & 56.5\%  &  39.4\%  &  4 & 69.4  &  74.1 \\
		69.2\%   & 61.5\%  &  40.8\%  &  7 & 72.3  &  73.2 \\
		66.2\%   & 57.3\%  &  38.8\%  &  7 & 69.8  &  72.4 \\
\hline
	\end{tabular}
	\label{tab:table1}
\end{table}

\begin{figure}
 \centering
  \includegraphics[scale=0.4]{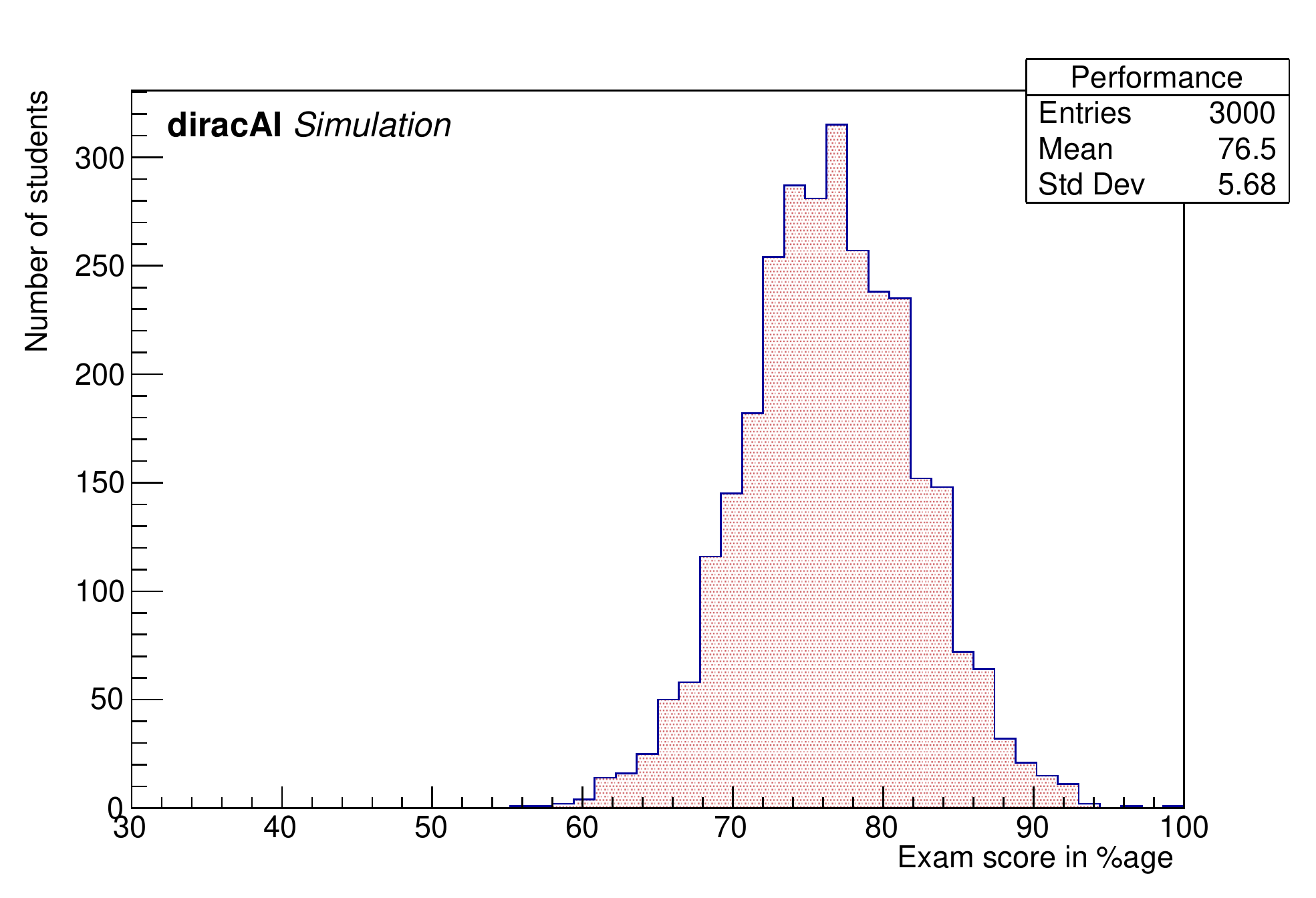}
  \caption{Exam performance }
  \label{fig:birds6}
\end{figure}

\section{Analysis }

 The  goal of the study is to demonstrate the concept for usage in a real-world scenario analysis. The objective is to model the outcome (performance) as a function of the input parameters. Since we are considering regression for this, the output will be modeled as a weighted linear combination of input variables similar to eqn. 4. We describe briefly the ML algorithm used for this study.
 
\subsection{Machine Learning Algorithm}

There exists several machine learning algorithms to model the data. Those are namely regression~\cite{schneider2010linear}, boosted decision trees (BDT)~\cite{ke2017lightgbm}, Xgboost~\cite{chen2016xgboost}, deep neural networks (DNN)~\cite{chien2018source}, support vector machines (SVM)~\cite{hearst1998support} etc. We have chosen regression for this study as it is comparatively easier and straight forward to interpret the results. A brief description of theory of regression is given below. 

\subsubsection{Linear Regression }

Linear regression is a supervised machine learning algorithm. It is a technique for investigating the relationship between independent variables or features and a dependent variable or outcome. It’s being used as a method for predictive modeling in machine learning, in which an algorithm is used to predict continuous (infinite) outcomes. We try to model the output (performance) assuming a linear combination of input parameters considered in this study. The hypothesis assumed in the linear regression can be written as the following for our case

\begin{equation}
H_{\theta}(x) = {\theta_{0} +\theta_{1}X^{1}+ \theta_{2}X^{2} + \theta_{3}X^{3} + \theta_{4}X^{4} + \theta_{5}X^{5}}
\end{equation}

where $\theta_{i}$ s are the fitted parameters to be determined. $X^{i}$s represent the input data points. We use a least square sum~\cite{miller2006method} as cost function \cite{shephard2012cost} for the study. The cost function is given by

\begin{equation}
J(\theta_{0},\theta_{1},\theta_{2},\theta_{3},\theta_{4},\theta_{5}) = {\frac{1}{2m} \sum_{1}^{m}(H(x^{i}) - y^{i})^{2}}
\end{equation}

where $y^{i}$ denotes the performance value. m is the number of training examples considered in modeling the data. $\mathbf{Gradient   }$ $\mathbf{descent }$ algorithm is used to optimize the cost function. The $\theta$s  obtained iteratively by the gradient descent algorithm are given by

\begin{equation}
{\theta_{j}} := {\theta_{j} - \alpha \frac{\partial}{\partial \theta_{j}} J(\theta_{0},\theta_{1},\theta_{2},\theta_{3},\theta_{4},\theta_{5}) }
\end{equation}

\subsection{Training }

For this study, we used simulated data with a total of 3000 points. To be used for training and testing, the entire data set is divided into two sets with a 4:1 ratio. A total of 2400 data points are used in the training and remaining 600 data points are used for testing.

\subsubsection{Learning Rate }
Learning rate is an important parameter that needs t be optimized to reduce the computing cost. But in the scope of current study, no such optimization is performed. The number of iterations are taken to be high to ensure the model converges. The set of parameters used in the training could be found in the following table. 

\begin{table}[h!]
	\caption{Parameters used in the training}
	\centering
	\begin{tabular}{ll}
	\hline
$\alpha$ & $\rm m_{train}$\\
\hline 
0.05 & 2400 \\
\hline
	\end{tabular}
	\label{tab:table2}
\end{table}

\subsubsection{Cost vs Number of iterations}

To ensure that the model does not under fit the data, we plot the value of cost versus the number of iterations. The number of iterations are taken enough to ensure the cost decreases and reaches a plateu. The cost vs. number of iterations graph is shown is Fig. 4.
\begin{figure}[h!]
  \centering
  \includegraphics[scale=0.38]{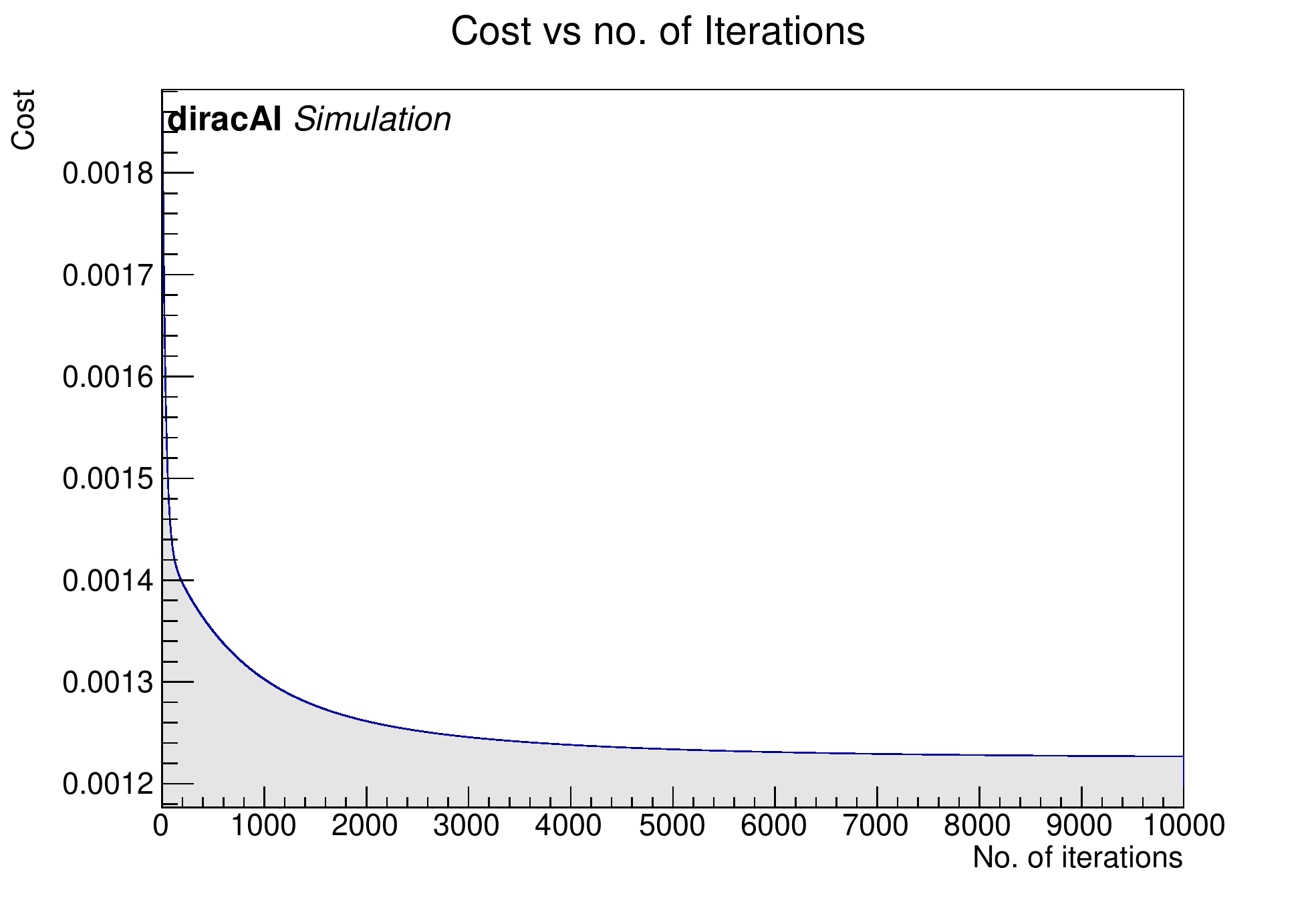}
  \caption{Cost vs number of iterations }
  \label{fig:birds7}
\end{figure}
\newpage
\subsubsection{Fitted theta parameters in train and test set}

The injected (used in data generation), train, test set $\theta_{i}$s are shown in the table.  It shows the injected values of $\theta$s and fitted $\theta$s are quite close and hence ensures that models works for the set of data considered in the study. The same model when applied to test data also retrieves the value of the $\theta$s injected to the simulated data set. Since simulated data is used here, a comparison with injected values of $\theta$s are possible in the scope of this study. The small differences between different sets could possibly further be improved by choosing a optimized set of $\alpha$ and number of iterations.

\begin{table}[h!]
	\caption{Trained $\theta$ parameters}
	\centering
	\begin{tabular}{lllllll}
	\hline
	Parameter &	$\theta_{0}$   &  $\theta_{1}$  &  $\theta_{2}$   & $\theta_{3}$ & $\theta_{4}$     & $\theta_{5}$ \\
	\hline
	Injected &	0.20   & 0.30 & 0.05   & 0.40 & 0.10   & 0.15 \\
    Fitted(training) &	0.19   & 0.23 & 0.03   & 0.43 & 0.12   & 0.20 \\
	Fitted(testing)  & 0.25 & 0.17 & 0.06 & 0.42 & 0.11 & 0.15\\ 
	\hline
	\end{tabular}
	\label{tab:table3}
\end{table}

\subsection{Interpretation of fitted parameters  }
 
 Once the fitted parameters are obtained from training, the model could be used for prediction. In the context of this study, $\theta_{i}$s are to be interpreted as the probabilities. As a result, the attendance coefficient ($\theta_{1}$) would show how significant the impact of attendance on final test achievement is. The same interpretation applies to all other parameters ($\theta_{i}$s). The teacher or institute owner wants to make sure that the obligations for the parameters with a high impact are met before those for those with a low impact.
 
\subsection{Credit Score }
The fitted values of the $\theta_{i}$s could be used to analyse students' exam weaknesses and create a single parameter known as a "credit score" to track students' daily development. The following equation illustrates the functional form of the credit score. 

\begin{equation}
\rm Credit~Score = {\theta_{0} +\theta_{1}X^{1}+ \theta_{2}X^{2} + \theta_{3}X^{3} + \theta_{4}X^{4} + \theta_{5}X^{5}}
\end{equation}

The $X^{i}$s denote the same meanings as in the equation 3. The values of $X^{i}$s are need to calculate the credit score for a student. In order to make sure the student will perform well in the exam, teacher have to make sure the credit score for every class is higher consistently. The class-credit-score of a student could be calculated using the values of $X^{i}$s collected from each class.

\section{Summary and Conclusion}
This study provides a proof-of-concept for how curricular activity data might be combined to create a single metric known as a credit score, which can then be used to analyse exam weaknesses and keep tabs on everyday activities. Throughout this study, we have used simulated data and several simplified scenarios to illustrate the concept. Although the study's input data do not have complex correlations with performance, real-world data may have more intricate structures, making it difficult to analyse the input data using regression. The input data could be modelled using advanced ML methods. Additionally, the uncertainties related to the final model are not attempted to be calculated in the context of the current study. The consistency of the model would be improved by taking extra care to comprehend and estimate the uncertainties related to the model. However, any real-world situation could benefit from the concept and methods used in this paper.\\

\textbf{\Large About the authors:}\\

{\bf \href{https://diracai.com/dr-bibhuprasad-mahakud/}{Dr Bibhuprasad Mahakud}} is the director and CEO of the \href {https://diracai.com}{diracAI Private Limited, Bhubaneswar, India}. Dr. Mahakud has completed his Masters at IIT, Delhi and Ph.D. in Experimental High Energy Physics at TIFR, Mumbai. He was a postdoctoral researcher at Purdue University, USA and at IIT, Bhubaneswar respectively. He has been a member of CMS Collaboration, CERN, Switzerland for $\sim$10 years. He was a visiting scientist at Fermilab, USA and was a visitor at CERN, Switzerland during his Ph.D. and postdoctoral periods. He is a co-author of more than 10 research publications of CMS Collaboration, CERN.\\

{\bf \href{https://diracai.com/dr-ipsit-panda/}{Mr. Ipsit Panda}} is the director of the \href {https://diracai.com}{diracAI Private Limited, Bhubaneswar, India}. Mr. Panda has completed his Masters at IIT, Delhi and has more than 10 years of teaching experience to the k12 students of India. He has made one research publication on his work on theoretical physics.\\

{\bf \href{https://amity.edu/faculty-detail.aspx?facultyID=4159}{Dr. Bibhuti Parida}} is a researcher and assistant professor of High Energy Physics at Amity University, Noida. Dr. Parida has completed his Masters and M.Phil. from Sambalpur University, Jyoti Vihar, Sambalpur, Odisha and M.Tech. from Tezpur Central University, Assam. He has performed his doctoral research work in the field of Experimental High Energy Physics at TIFR, Mumbai and postdoctoral research at Shanghai Jiao Tong University, China, at Tomsk State University, Russia and at Ben-Gurion University of the Negev, Israel respectively. He has worked as a visiting scientist at Fermilab, USA and has been a regular visitor to CERN, Switzerland during his PhD and postdoctoral stints. He is a co-author of more than 800 research publications of CMS and ATLAS Collaborations, CERN, Switzerland.\\

{\bf Declaration:}  The authors declare that they have no conflict of interest.

\bibliographystyle{unsrt}
\bibliography{bibliography.bib}

\end{document}